\documentclass{article}

\usepackage[margin=0.7in]{geometry}
\usepackage{amsmath}
\usepackage{amssymb}
\usepackage{slashed}
\usepackage{graphicx}
\usepackage{graphics}
\usepackage{epsfig}
\usepackage{color}
\usepackage{xcolor}
\usepackage{soul}
\usepackage{hyperref}
\usepackage{upgreek}
\usepackage{authblk}
\usepackage{blindtext}
\usepackage[section]{placeins}
\usepackage{mathrsfs}
\usepackage{dsfont}
\usepackage[normalem]{ulem}

\newcommand{\be}{\begin{equation}}
\newcommand{\ee}{\end{equation}}
\newcommand{\bea}{\begin{eqnarray}}
\newcommand{\eea}{\end{eqnarrray}}

\author[1,2]{Astrid Eichhorn \thanks{eichhorn@cp3.sdu.dk}}

\affil[1]{\textit{CP3-Origins, University of Southern Denmark, Campusvej 55, 5230 Odense M, Denmark}}
\affil[2]{\textit{{Institut f\"ur Theoretische
  Physik, Universit\"at Heidelberg, Philosophenweg 16, 69120
  Heidelberg, Germany}}}

\title{Asymptotically safe gravity\footnote{Presented at the 57th Course of the Erice International School of Subnuclear Physics, ``In search for the unexpected", June 2019}}

\date{}
\begin{document}

\baselineskip=0.5cm

\maketitle
\begin{abstract}  
This is an introduction to asymptotically safe quantum gravity, explaining the main idea of asymptotic safety and how it could solve the problem of predictivity in quantum gravity. In the first part, the concept of an asymptotically safe fixed point is discussed within the functional Renormalization Group framework for gravity, which is also briefly reviewed. A concise overview of key results on asymptotically safe gravity is followed by a short discussion of important open questions. The second part highlights how the interplay with matter provides observational consistency tests for all quantum-gravity models, followed by an overview of the state of results on asymptotic safety and its implications in gravity-matter models. Finally, effective asymptotic safety is briefly discussed as a scenario in which asymptotically safe gravity could be connected to other approaches to quantum gravity. 
\end{abstract} 

\section{Motivating quantum gravity}
On can start to think about quantum gravity from a seemingly simple question, namely: How do elementary particles interact via the gravitational field? We understand the answer to the analogous question for large bodies like stars, planets and satellites. It is encoded in General Relativity, which has been tested for these scales in multiple ways \cite{Berti:2015itd}. The Einstein equations encode the response of the spacetime geometry to the presence of energy and momentum,
\be
R_{\mu \nu} - \frac{1}{2}g_{\mu \nu}R = 8 \pi\, G_N(0) T_{\mu\nu}.
\ee
(We work in units where $\hbar = 1 =c$ and the reason for the unusual notation $G_N(0)$ for the Newton constant will become clear below.) Yet, there is no agreed upon treatment of quantum effects. For instance, how does the gravitational field of a particle in a superposition of position eigenstates look like? To answer this type of question, we need an understanding of quantum properties of spacetime geometry.

Further motivation for quantum gravity comes from the breakdown of 
the Standard Model (SM) of particle physics and General Relativity (GR), respectively. Both are incomplete, as we can infer from the existence of singularities in these theories. For GR, these are curvature singularities in black-hole spacetimes as well as cosmological spacetimes, which signal a breakdown of GR. This implies that we do actually not understand the true nature of the objects that are being observed indirectly through their gravitational-wave emission \cite{Abbott:2016blz} and imaged by the Event Horizon Telescope \cite{Akiyama:2019cqa}. The resolution of singularities is expected to come from quantum effects, motivating the search for a quantum theory of gravity. On dimensional grounds, such a theory is expected to become relevant at the Planck scale $M_{\rm Planck} =\sqrt{\hbar\, c\, G_N^{-1}(0)} \approx 10^{19}\, \rm GeV$.\\
 For the SM, the singularities occur in a subset of the interactions: Due to the screening nature of quantum fluctuations of a subset of SM-fields, some of the SM-couplings are driven towards increasing values as we ``zoom in", i.e., go towards higher energies. Within perturbation theory, they hit singularities, so-called Landau poles, at a finite energy, signalling the need for new physics
\footnote{The existence of Landau poles indicates the breakdown of perturbation theory, but not necessarily a breakdown of the theory which can continue to make physical sense at a nonperturbative level, just as QCD which features a perturbative Landau pole where it enters a nonperturbative regime. Nonperturbative studies of scalar theory as well as Quantum Electrodynamics indicate that there is no non-perturbative ultraviolet completion. In order to remove the ultraviolet lattice cutoff, the low-energy value of the corresponding couplings has to be chosen equal to zero, implying that the theory is noninteracting (trivial). It is generally assumed that the triviality problem will be inherited by the full SM.}.
A hint for the type of new physics that is missing comes from the scale of the Landau poles. Since the discovery of the Higgs particle at a mass of about 125 GeV \cite{Aad:2012tfa,Chatrchyan:2012xdj}, we know that these Landau poles lie well above the Planck scale. Thus, the LHC results actually imply that the SM is actually internally consistent up to (and beyond) the scale at which the onset of quantum gravity is expected \footnote{There are observational indications that the SM needs to be supplemented by some additional fields, among others to explain the dark matter in the universe. Depending on the dark-matter model, Landau poles can occur below the Planck scale, indicating the need for further new physics below that scale. On the other hand, there are dark-matter models that remain internally consistent up to the Planck scale and beyond. Further, it is not ruled out that dark matter consists of "known" physics, e.g., primordial black holes.}, motivating the minimalistic idea that an ultraviolet completion could arise through the inclusion of quantum gravity in the SM.

\section{The predictivity problem of perturbative quantum gravity}
A  conservative approach to quantum gravity is a quantum field theory of the metric, which does not introduce new fields for gravity and relies on the framework that has proven to be tremendously successful for all other fundamental interactions. Similarly to the photon for the electromagnetic field, or the gluon for the strong interactions, the quantum of the gravitational field is the graviton, a massless boson. Both photon and graviton mediate long-range interactions, while both gluons and gravitons feature self-interactions. The gauge symmetry at the heart of general relativity, diffeomorphism symmetry, removes all but the two helicity modes that a massless spin-2 field propagates \cite{vanderBij:1981ym}.
\\
In a quantum field theory, the key object to understand is the generating functional for the correlation functions, 
\be
Z = \int_{\Lambda_{\rm UV}} \mathcal{D}g_{\mu\nu}\,e^{iS[g_{\mu\nu}]},
\ee
where all field configurations of the metric field $g_{\mu\nu}$ (modulo diffeomorphisms) contribute and where the subscript $\Lambda_{\rm UV}$ indicates an ultraviolet cutoff \footnote{In other QFTs, the introduction of an ultraviolet cutoff is straightforward, as there is a background metric (typically the flat metric) that allows to define what one means by a UV mode. For quantum gravity, introducing a UV cutoff requires the introduction of an (auxiliary) background metric. In perturbation theory, one typically uses the flat metric.}. In this framework, the key question is not whether such a description is possible -- we already know that it is, using the effective field theory framework \cite{Donoghue:2017pgk} with a finite cutoff.
The key question instead is whether such a description can be  ultraviolet complete while also being predictive. To elucidate why predictivity is a nontrivial question, even though classical GR just has two free parameters (the Newton constant and the cosmological constant), let us review the perturbative quantization of gravity based on the Einstein-Hilbert action. Starting with the action
\be
S=\frac{1}{16\pi G_N}\int d^4x\sqrt{-g} R,\label{eq:EHact}
\ee
one expands
\be
g_{\mu\nu} = \eta_{\mu\nu} + \sqrt{\kappa}\, h_{\mu\nu},
\ee
where $\kappa = 8 \pi G_N$. Here, $\eta_{\mu\nu}$ is the Minkowski metric and $h_{\mu\nu}$ is the spin-2-field, the massless excitations of which are the two modes of the graviton. $\sqrt{-g}R$ actually contains arbitrarily high powers of $h_{\mu\nu}$ as well as two derivatives. Hence the propagator for $h_{\mu\nu}$ (which one obtains from the second-order term, supplemented by a suitable gauge-fixing) is quadratic in the momentum, as are all vertices (which one obtains from the higher-order terms). Therefore the expected highest power of divergence at loop order $L$ in $d$ spacetime dimensions is
\be
D = L(d-2)+2.
\ee
Next, we find out whether the expected divergences can be absorbed in the parameter of the Lagrangian --  the Newton coupling. To generate the highest divergence at $L$th loop order and $n$th order in the expansion in $h_{\mu\nu}$, the momenta at all vertices must be loop momenta. Therefore the term that is being generated is a momentum-independent $n$-graviton interaction. By diffeomorphism symmetry, this term must correspond to one that is generated from the expansion of the cosmological-constant term in powers of $h_{\mu\nu}$, as there is no  momentum-independent diffeomorphism invariant term except $\int d^4x\,\sqrt{-g}$. Thus we conclude that Eq.~\eqref{eq:EHact} should be supplemented by a cosmological-constant term.
The next, less divergent term that is generated depends on two external momenta, i.e., it is a two-derivative term (there cannot be terms with uneven numbers of derivatives as there is no diffeomorphism invariant way of contracting them to a scalar). Yet, this is exactly the type of term that is generated from the expansion of $\sqrt{-g}R$, i.e., the divergence can be absorbed in $G_N$. At one-loop order, this leaves us with the logarithmic divergence not yet taken care of. As it consists of four-derivative structures, this divergence cannot be absorbed in the parameters in the original Lagrangian, and requires curvature-squared terms \cite{tHooft:1974toh} \footnote{In four dimensions, the three diffeomorphism invariant four-derivative terms, $\int d^4x\,\sqrt{g}R^2$, $\int d^4x\,\sqrt{g}R_{\mu\nu}R^{\mu\nu}$ and $\int d^4x\,\sqrt{g}R_{\mu\nu\kappa \lambda}R^{\mu\nu\kappa \lambda}$ are not independent, as one particular combination of these corresponds to a topological invariant, the Gauss-Bonnet invariant. Further, in the absence of matter, the remaining two terms, $\int d^4x\,\sqrt{g}R^2$ and $\int d^4x\,\sqrt{g}R_{\mu\nu}R^{\mu\nu}$, vanish on shell. This makes quantum gravity in four dimensions in the absence of matter perturbatively renormalizable at one loop -- in a somewhat accidental fashion, not due to some underlying symmetry principle, as is the case in supergravity \cite{Bern:2018jmv}.}. The same pattern exists to higher loop orders \cite{Goroff:1985sz}: The logarithmic divergences require the addition of new terms to the Lagrangian in order to absorb the divergence. Each new term comes with a coupling, the low-energy value of which is a free parameter of the theory and needs to be fixed from observations. Therefore, at arbitrary high loop order, a perturbative quantization of Einstein-Hilbert gravity requires an infinite number of free parameters, rendering the theory non-predictive at high energies \footnote{At energies sufficiently beneath the Planck scale, where quantum-gravity effects are expected to be tiny, the effect of the higher-order terms is suppressed by positive powers of the energy over the Planck scale \cite{Donoghue:2017pgk}, thus the absence of predictivity does not pose a problem before one reaches the Planck scale.}.
 
To this observation, one could react in several distinct ways, covered, e.g., in \cite{Oriti:2009zz}, which have actually not been shown to necessarily be incompatible with each other, cf.~Sec.~\ref{sec:relntoapproaches}.
\begin{itemize}
\item One might conclude that perturbation theory fails for quantum gravity and instead pursue a nonperturbative quantization, as, e.g.,  in Loop Quantum Gravity.
\item One might conclude that the problem are transplanckian momentum modes, and therefore give up on local QFT beyond the Planck scale, as, e.g., in string theory as well as causal sets.
\item One might wonder whether there is a symmetry principle that one can impose that will generate relations between the infinitely many different couplings, such that only a finite number of free parameters remains, as in supergravity or asymptotically safe gravity.
\end{itemize}

\section{A first discussion of asymptotic safety}
Asymptotic safety is a quantum realization of scale symmetry. Scale symmetry relates physics at different scales. In classical field theory, the absence of dimensionful couplings is enough to ensure that the dynamics is scale-symmetric, i.e., does not depend on the scale. In quantum field theory, quantum fluctuations lead to a scale-anomaly, i.e., they break the classical symmetry. This is sourced by a scale-dependence of the couplings. Intuitively, it arises since quantum fluctuations turn the vacuum into a screening or antiscreening (depending on the type of interaction) medium, such that the value of the interaction strength depends on the scale. Therefore, it is not guaranteed that a theory that is consistent at one scale remains consistent as we change the scale to a smaller distance scale/larger momentum scale. In particular, divergences in the couplings can appear and signal the breakdown of the model. A restoration of scale symmetry is a way to avoid this and extend the theory up to arbitrarily short distance scales. This can be achieved if the effect of quantum fluctuations vanishes asymptotically, as in the case of asymptotic freedom, where the microscopic theory features classical scale symmetry. A genuine quantum realization of scale symmetry \cite{Wetterich:2019qzx} can be achieved in theories where the effect of quantum fluctuations balances out at finite values of couplings.
In this way, {\bf quantum scale symmetry} allows to construct models which hold up to arbitrarily short distance scales.
The more important aspect of quantum scale symmetry is the {\bf predictivity} it entails. One can think of quantum scale symmetry as just another symmetry one imposes on the dynamics of the theory. In general, symmetries restrict the possible interaction structures, thereby reducing the number of undetermined couplings, i.e., free parameters, of the model. In a similar way, imposing quantum scale symmetry entails relations between the couplings of a theory and thereby enhances the predictivity compared to the effective field theory setting, reducing the number of free parameters to a finite one.

Various examples of asymptotically safe quantum field theories exist, including perturbative gauge-Yukawa theories in four dimensions \cite{Litim:2014uca}; for more examples, see, e.g., the review \cite{Eichhorn:2018yfc}.

\section{Asymptotic safety with the functional Renormalization Group}
The idea of asymptotic safety can be understood in a Wilsonian way of thinking about path integrals. In particular, we will discuss the functional Renormalization Group (FRG) incarnation of the Wilsonian idea, pioneered for quantum gravity by Martin Reuter \cite{Reuter:1996cp}. 
Before we do this, let us stress that the idea of asymptotic safety and the framework  of the FRG are independent of each other. The FRG is simply a particularly well-suited tool to study this idea. In addition, lattice techniques in the form of Euclidean \cite{Laiho:2016nlp} and Causal Dynamical Triangulations \cite{Loll:2019rdj} as well as Regge calculus \cite{Hamber:2015jja} and tensor models \cite{Eichhorn:2018phj} are used to search for asymptotic safety in gravity. The very first studies of asymptotic safety were done within perturbation theory around 2 dimensions: For $d = 2+ \epsilon$, indications for asymptotic safety were found in the first two orders in the $\epsilon$ expansion, see, e.g., \cite{Gastmans:1977ad,Christensen:1978sc,Aida:1996zn}. Perturbation theory in $d=4$ was used to study asymptotic safety in \cite{Niedermaier:2009zz}.
\\
Note that from here on we work in Riemannian signature $(+,+,+,+)$, which is of course not the physically realized signature of the metric. The reason is purely technical -- we will introduce a momentum cutoff, and the Lorentzian momentum $p^2 = p_0^2-p_i^2$ cannot be restricted such that $p^2<k^2$ actually means that one does not have short (spatial) wavelengths present. In a QFT on a flat background, the relation between the Euclidean and the Minkowskian QFT is given by a Wick-rotation. In a quantum gravitational context one does in general not expect the Wick-rotation to exist \cite{Baldazzi:2018mtl}, and the relation between Lorentzian and Riemannian quantum gravity is unclear\footnote{The configuration spaces for Riemannian and Lorentzian metrics also differ in their global properties, see \cite{Demmel:2015zfa}.}. One should therefore keep in mind that from now on we explore a statistical theory (since the measure will be $e^{-S}$) of space (where configurations have  $(+,+,+,+)$ signature) instead of a quantum theory (with a quantum phase $e^{iS}$) of spacetime (where configurations have  Lorentzian signature).

\subsection{The functional Renormalization Group framework}
Unlike in standard perturbation theory, where one adds quantum fluctuations order by order in a loop expansion, but performs the integral over all momenta at once, in a Wilsonian setting one instead decomposes the path integral into "momentum-shells". For a QFT on a flat background, a momentum-shell contains all those field configurations with four-momentum squared $p^2$ in a range $\delta$ around some momentum $k$. 
The first ingredient we need for the decomposition of the gravitational path integral into momentum shells, is a background metric $\bar{g}_{\mu \nu}$. The corresponding covariant Laplacian $-\bar{g}^{\mu\nu}\bar{D}_{\mu}\bar{D}_{\nu}$ has eigenvalues $\lambda_s$ (they depend on the spin $s$ of the field they act on since the connection-part of the covariant derivative $\bar{D}_{\mu}$ is sensitive to the spin), which becomes four-momentum squared if one chooses $\bar{g}_{\mu\nu}= \delta_{\mu\nu}$. Given a momentum scale $k$, the eigenvalues can be sorted into UV modes ($\lambda_s>k^2$) and IR modes ($\lambda_s < k^2$). The background metric allows us to introduce a cutoff on the modes. Coordinate transformations, 
\be
\delta \gamma_{\mu\nu} = \mathcal{L}_v \gamma_{\mu\nu}=v^{\rho}\partial_{\rho}\gamma_{\mu\nu}+ (\partial_{\mu}v^{\rho})\gamma_{\rho_{\nu}}+ (\partial_{\nu}v^{\rho})\gamma_{\mu\rho},
\ee
relate field configurations with various eigenvalues $\lambda_s$, therefore the cutoff breaks diffeomorphism symmetry for the metric $\gamma_{\mu\nu}$ (the integration variable in the path integral). Nevertheless, we can preserve an auxiliary background diffeomorphism symmetry (under which also $\delta \bar{g}_{\mu\nu}= \mathcal{L}_v \bar{g}_{\mu\nu}$) -- as is standard in the background field approach, see also \cite{Reuter:1996cp}. It is most convenient to write the ``mode-sorting term" (which is actually called the regulator term) in a quadratic form, i.e., a mass-like term (this also makes it obvious that it breaks the diffeomorphism symmetry for the metric $\gamma_{\mu\nu}$). The quadratic nature of the regulator term is important, as it entails that the functional differential equation one can derive has a one-loop structure (not in the sense of perturbation theory, just in the sense of the number of momentum integrals) and therefore admits practical calculations with relative ease. Thus we write the Euclidean generating functional in a $k$-dependent fashion,
\be
Z_k [J]= \int_{\Lambda_{\rm UV}} \mathcal{D}\gamma_{\mu\nu}\, e^{-S[\gamma_{\mu\nu}]+ \int d^4x\sqrt{\bar{g}} J^{\mu\nu}\gamma_{\mu\nu} - \frac{1}{2}\int d^4x\sqrt{\bar{g}}\, \gamma_{\mu \nu}R_k^{\mu\nu\kappa\lambda}(-\bar{g}^{\alpha\beta}\bar{D}_{\alpha}\bar{D}_{\beta})\gamma_{\kappa \lambda}},
\ee
where $J_{\mu\nu}$ is a source term.
At this stage we keep the UV cutoff $\Lambda_{\rm UV}$, as otherwise the path integral is ill-defined. 
We have used the notation $\gamma_{\mu\nu}$ for the integration variable in the path integral, because we will reserve $g_{\mu\nu}$ for the expectation value, $\langle \gamma_{\mu\nu}\rangle = g_{\mu\nu}$.
Next we define the effective average action, by using a modified Legendre transform
\be
\Gamma_k[\bar{g}_{\mu\nu}, g_{\mu\nu}] = \underset{J}{\rm sup}\left( \int_x J^{\mu\nu}\cdot g_{\mu\nu}- \ln Z_k[K]\right) - \frac{1}{2}\int d^4x\sqrt{\bar{g}}\, g_{\mu \nu}R_k^{\mu\nu\kappa\lambda}(-\bar{g}^{\alpha\beta}\bar{D}_{\alpha}\bar{D}_{\beta})g_{\kappa \lambda}.
\label{eq:Gammadef}
\ee
 At $k=0$, this definition reduces to the standard definition of the effective action which provides the equations of motion for the expectation values of the quantum fields, in complete analogy to how the classical action provides the equations of motion for the classical field, i.e., $\delta \Gamma_{k=0}[\bar{g},g=\bar{g}]/\delta \bar{g}_{\mu\nu}=0$.

 The key advantage of the definition \eqref{eq:Gammadef} is that there is an exact functional differential equation that the scale-derivative of $\Gamma_k$ satisfies \cite{Wetterich:1992yh,Morris:1993qb,Ellwanger:1993mw,Reuter:1993kw}:
\be
\partial_t \Gamma_k[\bar{g},g] = k\, \partial_k\Gamma_k = \frac{1}{2}{\rm Tr}\left(\Gamma_k^{(0,2)}[\bar{g}_{\mu\nu},g_{\mu\nu}]+R_k \right)^{-1}\,\partial_t R_k.\label{eq:floweq}
\ee
Herein we have suppressed the spacetime indices of all quantities, including the regulator. $\Gamma^{(2)}_k$
 denotes the second functional derivative with respect to the metric, 
 \be
\Gamma_k^{(0,2)}[\bar{g},g]_{\mu\nu\kappa\lambda} =\frac{1}{\sqrt{\bar{g}(x)\bar{g}(y)}}\frac{\delta^2\Gamma_k}{\delta g_{\mu\nu}(x)\delta g_{\kappa \lambda}(y)}
\ee
 The trace includes a trace over spacetime indices, internal indices (which are present as soon as we add matter fields with internal symmetries) and an integration over spacetime. In terms of $-\bar{D}^2$, the trace becomes a sum/integral over the discrete/continuous spectrum of this operator.

The flow equation \eqref{eq:floweq} has two main uses: Given an initial condition $\Gamma_{k=\Lambda} = S_{\Lambda}$, one can integrate it down to obtain the corresponding effective description of the macroscopic dynamics. This is the main use in many areas of high-energy physics as well as condensed matter and statistical physics. The second use is to search for points in the space of dynamics at which quantum scale invariance is realized, i.e., the scale-derivative of the dynamics vanishes. The thereby-defined points are Renormalization Group fixed points. RG trajectories that emanate from such a point define asymptotically safe theories.

\subsection{Asymptotic safety in the FRG framework}
The effective average action $\Gamma_k$ in $d$ dimensions can be expanded in terms of quasilocal operators which are multiplied by $k$-dependent, i.e., running couplings, e.g., for gravity
\be
\Gamma_k[\bar{g}, g] = \frac{1}{16\pi G_N(k)}\int d^dx\sqrt{g}\left(R- 2\Lambda_k \right) + \sum_{i=2}^{\infty}\bar{a}_i \int d^dx\sqrt{g}R^i + \dots, \label{eq:GammakEHandbeyond}
\ee
where the dots indicate further terms that can be constructed out of non-negative powers of the Ricci scalar, Ricci tensor and Riemann tensor and non-negative powers of covariant derivatives, and also include terms that depend on the background metric -- we will come back to these later. Note that here the term ``running coupling" refers to the dependence of the couplings on the Wilsonian cutoff scale $k$. This differs from the use of the term in much of the perturbative literature, where it refers to the logarithmic dependence of couplings on the physical momentum \footnote{One can expand the effective action in a form that clarifies where the physical momentum dependence is, by using form-factors, e.g., in \cite{Bosma:2019aiu,Knorr:2019atm}:
\be
\Gamma_k =  \int d^4x \sqrt{g} \left( R f_1 (-D^2) R + R_{\mu\nu}f_2(-D^2)R^{\mu\nu} + R_{\mu\nu\kappa \lambda}f_3(-D^2)R^{\mu\nu\kappa\lambda}+...  \right).\label{eq:formfactor}
\ee 
Alternatively, one can also use an expansion in terms of momentum-dependent vertices \cite{Denz:2016qks,Knorr:2017fus,Christiansen:2017bsy,Eichhorn:2018nda}, which more directly relates to scattering processes and their momentum dependence.}.

Asymptotic safety implies that the scale-dependence of all the dimensionless counterparts of the essential couplings (those that cannot be removed by a (quasilocal) field-redefinition) vanish. The dimensionless counterparts are defined by multiplication with an appropriate power of $k$, e.g., for a coupling $\bar{g}_i$ of canonical mass-dimension $d_{\bar{g}_i}$, the dimensionless counterpart is defined as
\be
g_i = \bar{g}_i\, k^{-d_{\bar{g}_i}}. \label{eq:dimlessgi}
\ee
For instance, for the Newton coupling, cosmological constant and couplings of the $i$th power of the curvature we have
\be
G(k) = G_N(k)\, k^{d-2}, \quad \lambda(k)= \Lambda\, k^{-2}, \quad a_i = \bar{a}_i\, k^{d-2i}.
\ee
Their dimensionless scale derivatives define the beta functions
\be
\beta_G = \partial_t\, G(k), \quad \beta_{\lambda} = \partial_t\, \lambda(k), \quad \dots
\ee
An RG fixed point is a point in theory space (the space of all essential couplings) at which all beta functions vanish,
\be
\beta_{g_i} =0 \,\,\, {\rm at} \, \,\, g_{j}= g_{j\, \ast}\, \, \forall i, j.
\ee 
For various mechanisms that can induce such a fixed point and corresponding examples in various non-gravitational theories, see, e.g., the overview in \cite{Eichhorn:2018yfc}.\\
Demanding that the theory is defined on an RG trajectory that emanates from the fixed point as one lowers $k$ towards the infrared i) provides an ultraviolet complete theory and ii) is expected to impose predictivity on the theory. This second point can be seen as follows, as also discussed by Weinberg in \cite{Weinberg:1976xy}: Linearizing the beta function $\beta_{g_i}$ of some coupling $g_i$ around the fixed point to first order results in a linear differential equation
\be
\beta_{g_i} = \beta_{g_i}\Big|_{\vec{g}= \vec{g}_{\ast}}+ \sum_j \frac{\partial \beta_{g_i}}{\partial g_j} \Big|_{\vec{g}= \vec{g}_{\ast}} \left(g_j - g_{j\, \ast} \right) = 0 +  \sum_j \frac{\partial \beta_{g_i}}{\partial g_j} \Big|_{\vec{g}= \vec{g}_{\ast}} \left(g_j - g_{j\, \ast} \right).
\ee
 that is solved straightforwardly:
\be
g_i(k) = g_{i\, \ast} + \sum_I C^I\, V^I_i \left(\frac{k}{k_0}\right)^{-\theta_I}.\label{eq:linearizedgi}
\ee
Herein $V^I$ are the eigenvectors of the stability matrix $ \partial \beta_{g_i}/(\partial g_j) \Big|_{\vec{g}= \vec{g}_{\ast}}$. The critical exponents $\theta_I$ are the eigenvalues of the stability matrix, multiplied by an additional negative sign. The reason for the sign will become clear below \footnote{Note that the opposite sign convention is sometimes used for the critical exponents.}.
The constants of integration, the $C^I$, are the free parameters of the theory. For an arbitrary trajectory that happens to enter the linearized regime close to the fixed point at some $k_0$, one requires knowledge on all the infinitely many $C^I$ in order to determine the trajectory exactly. Yet, the presence of the fixed point has an important effect, seen from Eq.~\eqref{eq:linearizedgi}: Lowering $k/k_0$, i.e., flowing towards the IR, one notices that the contribution from those eigenvectors $V^I$ which feature a negative critical exponent, $\theta_I<0$, is suppressed, such that the corresponding $C^I$ become irrelevant for the values of the couplings at $k_0\ll k$. Therefore the corresponding directions in the space of couplings are called irrelevant (which is the same as IR-attractive or UV-repulsive). Conversely, directions for which $\theta_I>0$, become relevant in the IR, since the associated term in Eq.~\eqref{eq:linearizedgi} grows towards the IR, such that the IR-values of the couplings depend on the value of $C^I$. The relevant directions are also called IR-repulsive (or UV attractive), since those are the directions in which the fixed point repulses the RG flow towards the IR.

\begin{figure}[!t]
\begin{minipage}{0.5\linewidth}
\includegraphics[width=0.8\linewidth,clip=true, trim=9cm 4cm 8cm 4cm]{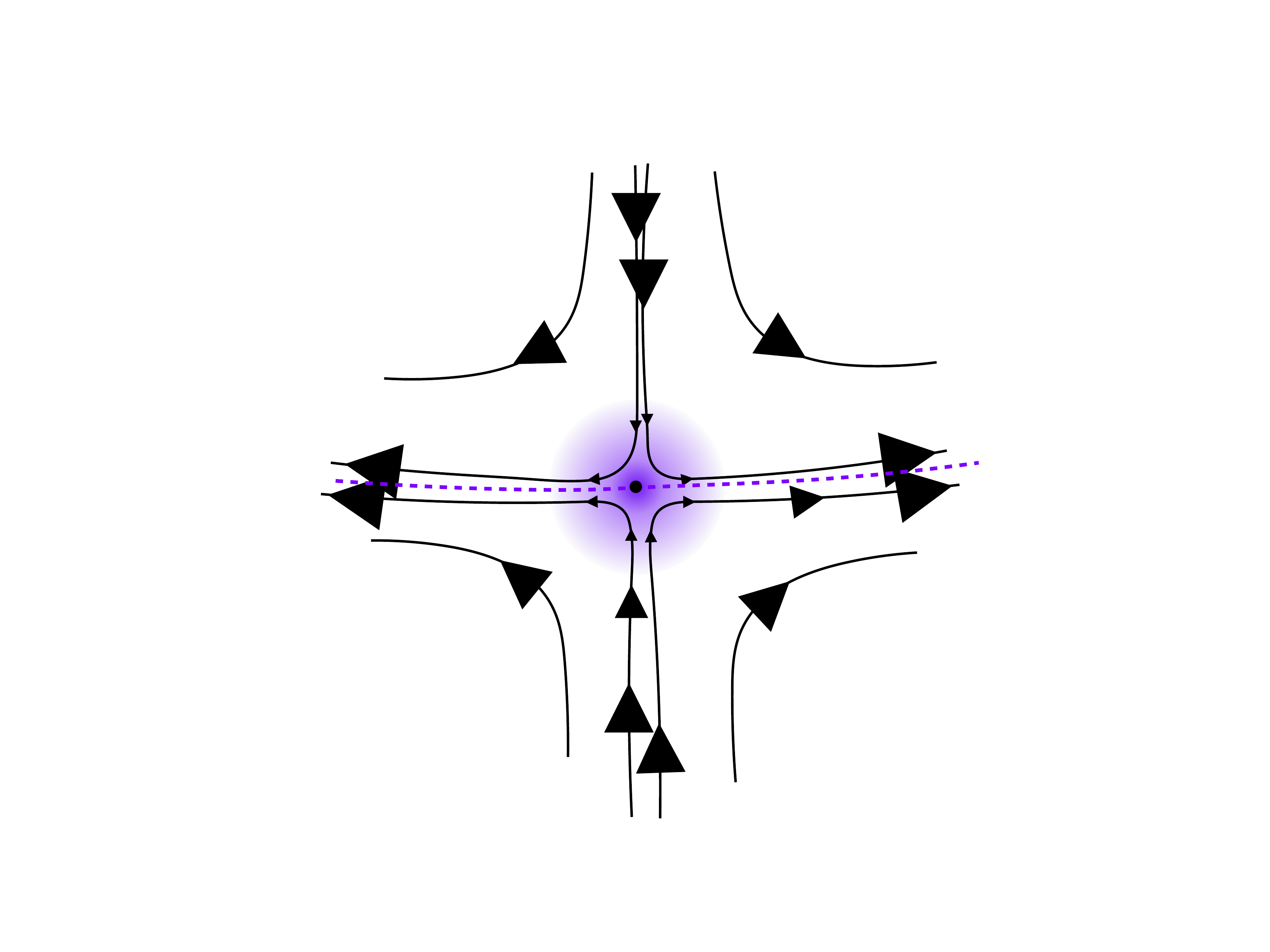}
\end{minipage}
\quad
\begin{minipage}{0.45\linewidth}
\caption{\label{fig:illFP} Illustration of a fixed point (black dot) with one relevant and one irrelevant direction. The critical hypersurface, to which trajectories are pulled towards the infrared, is indicated by the purple dashed line.
The speed of the flow is indicated by the arrows: In the vicinity of the fixed point, the flow slows down very significantly.}
\end{minipage}
\end{figure}

At any finite ratio $k/k_0$, a tiny contribution of the $C^I$ belonging to irrelevant directions remains in the values of all couplings. It shrinks when the ratio $k/k_0$ is increased further, i.e., the further the domain of validity of the quantum-field theoretic description extends, the more predictive it becomes.
Specifically, when one demands that the theory is UV complete, one shifts the scale at which the trajectory enters the linearized regime around the fixed point to infinity, $k_0 \rightarrow \infty$. Then, the IR values of the couplings on such a trajectory lose all ``memory" of those $C^I$ that belong to irrelevant directions, and the flow is confined to the so-called critical surface of the fixed point\footnote{Note that the discussion of how a fixed point imposes predictivity is sometimes phrased in terms of RG flows towards the UV, and in terms of UV-repulsive directions. While in practise one can invert the RG flow in finite-dimensional subspaces of the theory space and ``flow towards the UV", the actual direction of the RG flow is always from the UV to the IR, i.e., the microphysics determines the macrophysics, not vice-versa.}. Therefore, the IR values of the couplings, and thus the physics, on such a trajectory is purely determined by the relevant directions\footnote{A theory space can feature more than one fixed point. The critical surfaces of different fixed points differ. Therefore, one can distinguish the various fixed points in terms of their predictivity and ultimately in terms of the physical implications, i.e., the different effective actions at $k=0$.}. \\

What is it about a coupling that determines whether it is relevant or irrelevant?\\
 Let us first focus on the free fixed point. There, the diagonal entries of the stability matrix are determined by the canonical dimensions of the couplings, cf.~Eq.~\eqref{eq:dimlessgi}, since $\beta_{g_i}= -d_{\bar{g}_i}g_i+\mathcal{O}(\vec{g}^2)$, such that $\theta_I = d_{\bar{g}_i}$ (with the sign convention for the $\theta_I$ we use). Therefore, in the perturbative regime, the RG flow automatically drives higher-order (i.e., so-called perturbatively nonrenormalizable) couplings to zero, while those of positive canonical dimension are relevant. Dimensionless couplings are associated to vanishing critical exponents, and are either marginal (when their beta function vanishes to all orders), marginally relevant (when the leading term in their beta function is negative) or marginally irrelevant (when the leading term in their beta function is positive).\\
 To understand the case of an interacting fixed point, let us assume that there is a parameter that allows to deform beta functions in such a way that the free fixed point continuously deforms into an interacting one \footnote{For many systems, the dimensionality constitutes such a parameter, see, e.g., the list in \cite{Eichhorn:2018yfc} and references therein.}. Under this deformation, the relevant and irrelevant directions become associated to superpositions of the couplings instead of aligning with the axis of theory space. Further, the critical exponents can no longer be calculated just from the knowledge of the canonical dimensions, there is a finite contribution from higher-order terms in the beta function. This contribution can render canonically irrelevant couplings relevant and vice-versa. As long as the contribution is finite, only finitely many couplings are relevant at the fixed point, since there is still the canonical contribution from the dimensionality of the couplings which becomes increasingly irrelevant with higher order in derivatives and fields.\\

In summary, asymptotic safety is the property of the space of couplings to feature an interacting RG fixed point with a finite number of relevant directions. Since the scale $k$ is introduced into the path integral as an auxiliary parameter, the $k$-scaling is of course unphysical, so why does it matter whether there is such a fixed point? 

Firstly, it implies that the path integral is actually well-defined in the sense that one can take a continuum limit ($k$ can be viewed similarly to an inverse lattice spacing). Secondly, it results in a quantum field theory parameterized by a finite number of free parameters. It should be emphasized that these do not directly correspond to measurable quantities (couplings are in general not observables). Yet, they determine the full effective action at $k=0$, which in turn determines the correlation functions. Correlation functions of suitable gauge-invariant operators are observables \footnote{In quantum gravity, the definition of local observables is in general a thorny problem. For many actual physical situations, such as, e.g., in particle physics or cosmology, correlation functions of excitations around a flat or cosmological background become suitable observables.}, and therefore measurable. Asymptotic safety is phenomenologically viable, if the effective dynamics in the limit $k \rightarrow 0$ which is fully determined in both its local as well as non-local parts by the finite number of relevant directions, agrees with observations. Note that it is the existence of the fixed point which results in an effective dynamics with just a finite number of free parameters, whereas the effective action in an effective field theory in general features an infinite number of free parameters, and can therefore only be used for predictions in special cases where all but a finite number of operators become dynamically unimportant.

Secondly one expects that the existence of a scaling regime is a general property that will also affect the physical scale dependence, i.e., the presence of a scaling regime at high $k$ is expected to be mimicked by scaling properties at high momentum/curvature scales. Note that many physical situations exhibit more than just one scale, therefore the analysis of such scaling-regimes can require some care. A cautionary example that highlights that a direct replacement of $k$ by physical momentum scales can lead to wrong results is given in \cite{Anber:2011ut}. In particular, the example highlights that it is important to keep track of the various momentum channels that in general characterize scattering amplitudes, and that cannot always be captured by a simple scale-dependent coupling but in general require momentum-dependent vertex functions. Steps towards investigating actual physical momentum dependencies in asymptotically safe gravity have been done both for correlation functions in a vertex expansion, see, e.g., \cite{Denz:2016qks,Knorr:2017fus,Christiansen:2017bsy,Eichhorn:2018nda} as well as in a form-factor approach \cite{Bosma:2019aiu,Knorr:2019atm}, cf.~Eq.~\eqref{eq:formfactor}.

\subsection{Brief overview of key results in pure gravity}
In terms of an expansion of the effective action in quasilocal curvature operators, such as in Eq.~\eqref{eq:GammakEHandbeyond}, there is a large body of literature. In summary, it shows a fixed point in any finite truncation, including i) powers of the Ricci scalar up to $R^{70}$ \cite{Falls:2018ylp}, ii) a complete truncation to order curvature squared \cite{Benedetti:2009rx}, iii) Einstein-Hilbert plus the Goroff-Sagnotti curvature-cubed term \cite{Gies:2016con}. The study in \cite{Falls:2017lst} accounts for the effect of the different structure of the metric propagator derived from an expansion containing powers of the Ricci tensor. In these studies, not more than 3 couplings are relevant, and the critical exponents exhibit near-canonical scaling in those truncations that reach high orders in derivatives.
A more extended overview of the literature providing more details can be found, e.g., in \cite{Percacci:2017fkn,Reuter:2019byg,Eichhorn:2018yfc}.

Due to the fact that we introduced a {\bf local coarse-graining procedure}, the {\bf background metric} appears as an argument in the effective action, and accordingly background-couplings appear in the expansion of $\Gamma_k$. Their scale dependence differs from the dynamical couplings (those in Eq.~\eqref{eq:GammakEHandbeyond}). There are two sources for this difference: Firstly, the regulator term itself introduces a difference between the two fields. Secondly, since the flow equation is based on the propagator of metric fluctuations, a gauge-fixing term needs to be introduced, which can be done with the help of the background metric. Therefore, the theory space actually i) contains couplings of operators constructed from the dynamical metric as well as operators which are also constructed from the background metric ii) contains terms that violate the ``dynamical" diffeomorphism symmetry (under which the dynamical metric transforms), since we are in a gauge-fixed setting. There are symmetry-identities that relate some of these couplings:  a) Slavnov-Taylor identities constrain the gauge-variant correlation functions for the dynamical fields;  b) a (modified) shift identity relates the background couplings to the dynamical couplings, and accounts for the fact that background-independence is broken. The latter has been explicitly explored for quantum gravity in \cite{Dietz:2015owa,Labus:2016lkh,Percacci:2016arh,Ohta:2017dsq,Eichhorn:2018akn}.\\
As usual in the background field formalism for gauge theories, an (auxiliary) background diffeomorphism symmetry can be retained. This is crucial for $\Gamma_{k \rightarrow 0}[\bar{g}, \bar{g}]$ which encodes the physics of the theory. It is nevertheless crucial to keep track of the terms pertaining to the dynamical metric, as these drive the flow of the effective action. 
The distinction between the fluctuating metric $g_{\mu\nu}$ and the background metric $\bar{g}_{\mu\nu}$ is the subject of a growing number of works, see the review \cite{Reichert2020}. The truncations that have been explored in this setting so far feature a fixed point with very similar properties to that found under the simplifying approximation $\Gamma_k^{(0,2)}[\bar{g},g]= \Gamma_k^{(2,0)}[\bar{g},g]$, the single-metric approximation.

Setting up flows  in a ``bimetric" language with $\bar{g}_{\mu\nu}$ and $g_{\mu\nu}$, see \cite{Becker:2014qya} and references therein, is equivalent to a fluctuation-field setup, see \cite{Reichert2020} and references therein.
In terms of correlation functions for the fluctuation field $h_{\mu\nu} = \bar{g}_{\mu\nu}- g_{\mu\nu}$ in the linear split, work started with the momentum-dependence of the graviton \cite{Christiansen:2012rx} \footnote{For the anomalous dimensions of the Faddeev-Popov ghosts, see \cite{Groh:2010ta,Eichhorn:2010tb}.}. By now, momentum-dependent 3- and 4-point functions for metric fluctuations have been explored \cite{Christiansen:2015rva,Denz:2016qks}, showing a behavior that appears to be consistent with the onset of apparent convergence. These studies refer to a flat background; upgrades to a curved background can be found in \cite{Knorr:2017fus,Christiansen:2017bsy}.

\subsection{Key open questions in brief}
For a more detailed discussion of key open questions, see \cite{ASopenquestions}. In brief, the status of asymptotically safe gravity is that an interacting fixed point, called the Reuter fixed point, has been discovered in increasingly elaborate truncations which has given rise to the general expectation that the fixed point exists in full theory space. Nevertheless, {\bf quantitative apparent convergence} of the results has not yet been achieved, in particular in the presence of matter (see below).\\
As stressed above, the FRG setup is a Euclidean one, which implies that the question of whether asymptotic safety is realized in {\bf Lorentzian quantum gravity}, is an open one. An analytical continuation of the flow equation Eq.~\eqref{eq:floweq} for non-gravitational settings is discussed in \cite{Floerchinger:2011sc} and explored in a simple truncation for gravity in \cite{Manrique:2011jc}.\\
Further, the {\bf number and nature of propagating degrees of freedom} is an important open question. It is not uniquely determined by saying that the metric field carries the gravitational degrees of freedom, since the actual degrees of freedom depend on the metric propagator: Generically, a higher-order propagator \footnote{Since one is investigating a gauge field, one has to keep in mind that the usual discussion of the K\"allen-Lehmann representation of the propagator does not apply. Further, the propagator of metric fluctuations around a flat background is not necessarily the relevant object to investigate, if the vacuum of the theory differs from flat spacetime.} implies the existence of further propagating degrees of freedom, just as a higher-order classical equation of motion generically (but not always! \cite{Barnaby:2007ve}) implies the need for additional initial conditions, and thus degrees of freedom.  These additional degrees of freedom might be tachyonic and/or ghost modes. If they are ghosts and their masses are high enough, then ``effective asymptotic safety" can still hold, i.e., the RG trajectories have a finite UV cutoff, but can still spend a large ``amount" of RG time close to the fixed point, such that the predictive power of the fixed point is still (approximately) realized. A key question for asymptotic safety is the fate of ghost-like poles under extensions of the truncation: whereas any polynomial truncation to finite order in momenta is expected to exhibit these, it is unclear whether their masses can move to infinite values under extensions of the truncation, effectively decoupling these modes from the theory.\\
Regarding the physical implications of the theory, the {\bf resolution of classical spacetime singularities} is a key requirement. Here, asymptotic-safety inspired upgrades of black-hole spacetimes have been investigated and found to be singularity free \cite{Bonanno:2000ep,Platania:2019kyx}. It is key to stress that these spacetimes are not derived as solutions of the full dynamics of the quantum theory, instead they are obtained through an RG improvement procedure, and it is not known whether they actually capture the physics content of asymptotic safety. The question of the so-called information paradox depends on the final state of black-hole evaporation, which has been explored in the asymptotic-safety inspired spacetimes \cite{Bonanno:2006eu}.

\section{Observational tests of quantum gravity}
In physics, the connection of theoretical proposals to experimental and observational tests is key; internal self-consistency is an insufficient criterion to decide whether a theory describes nature and testability truly matters. How then do we test a quantum theory of gravity? 
One expects quantum gravity to have been important in the early universe, such that observable imprints might have been left in the spectrum of primordial tensor modes. A second environment with high curvature scales, at which terms beyond Einstein gravity, such as they typically arise in quantizations of gravity, could become important, are black- hole spacetimes. Yet, an inspection of the curvature scales currently being probed  \cite{Baker:2014zba} by, e.g., black hole binary mergers as well as the Event Horizon Telescope, reveals that the curvature scale at which quantum gravitational effects set in would have to be rather low in order to lead to detectable effects -- at least within a local treatment.

There is a different route towards observational tests of quantum gravity, which is not through direct tests, i.e., checks of \emph{new} effects, but by using consistency with already performed experimental tests. This concerns, e.g., the requirement to reproduce GR (implying, among others, the absence of large Lorentz-symmetry violations \cite{Gumrukcuoglu:2017ijh}). 
Consistency tests in the purely gravitational sector are actually just one possibility.
In fact, the existence of matter is critical for observational consistency tests of quantum gravity that are possible right now \footnote{These are possible from an observational point of view, i.e., the data is already there. The corresponding predictions for this data from a ``fundamental" model including quantum gravity is mostly lacking, and is a theoretical challenge for quantum-gravity practitioners. Indeed, the situation is not, as is so often stated, that there is a lack of experimental data to test quantum gravity; the data is there, but the comparison with theoretical predictions is challenging due to the technical complexity of the problem.}.
Indeed, requiring that the low-energy limit of some ``fundamental" model includes all Standard-Model particles (but no unobserved extra ones) with the appropriate quantum numbers and interaction strengths provides  observational consistency tests that could rule out quantum-gravity models without the need for direct tests of Planck-scale physics \footnote{The ``fundamental" model can in principle be an explicit unification of gravity and particle physics, as attempted in string theory, a joint asymptotically safe model of the metric field and all matter fields, as in asymptotic safety, or a description from which both gravitational as well as matter degrees of freedom ``emerge". In general one might expect that the more structure one requires to be ``emergent", the more difficult this feat will be to achieve. In this sense, asymptotic safety might be the most conservative framework to try: One simply aims to extend a description that already works well over a larger range of scales.}.

\subsection{Towards observational consistency tests from the interplay with matter}
To be concrete, we will discuss a number of tests and the status of the answer within asymptotically safe gravity. In general, none of the answers are definite yet, as they are all given within truncations of Euclidean quantum gravity. Thus, there is a systematic uncertainty attached to the approximation in addition to the question whether calculations within a Euclidean quantum-gravity context can indeed be meaningfully compared to observations.

\subsubsection{Number of matter fields}
Accommodating the correct number of matter fields is a first nontrivial requirement. As a motivating analogy of why this is a nontrivial test, consider SU($N_c$) Yang-Mills theory with $N_f$ fermions in the fundamental representation of the gauge group. Due to the screening nature of fundamental fermions, asymptotic freedom is lost beyond a critical value of $N_f$. In a similar manner, the impact of quantum fluctuations of scalars, fermions and vectors might support or prevent asymptotic safety in gravity.
Minimally coupling these fields to the Einstein-Hilbert and $f(R)$ truncations results in the possibility to consistently include all Standard-Model fields \cite{Dona:2013qba,Meibohm:2015twa,Biemans:2017zca,Alkofer:2018fxj,Wetterich:2019zdo}, see also \cite{Eichhorn:2015bna} for the unimodular case. \\
As in pure-gravity systems, background couplings and dynamical couplings differ. This difference has been tackled by exploring $n$-point fluctuation field correlation functions for gravity-matter systems for $n=2,3$.  Within the corresponding truncations, the Reuter fixed point exists and can be continued to finite numbers of matter fields, see \cite{Meibohm:2015twa,Dona:2015tnf,Eichhorn:2018ydy,Eichhorn:2018nda}.\\
Let us point out a key limitation of these studies:
Within asymptotically safe gravity, the interacting nature of gravity percolates into the matter sector and results in the necessary presence of matter self-interactions, see, e.g., \cite{Eichhorn:2011pc,Eichhorn:2012va}. In studies exploring whether there is a bound on the number of matter fields compatible with asymptotic safety in gravity, truncations to date set the corresponding couplings to zero. In particular at large numbers of matter fields, the ``backreaction" effect of these induced interactions could matter. 

\subsubsection{Existence of light fermions}
Fermions in the Standard Model are light compared to the Planck scale. This is a consequence of the fact that their masses are set by Yukawa couplings which are perturbatively small, and Dirac mass terms are absent.

A system of $N_f$ fermions with a global ${\rm SU}(N_f)_L \times {\rm SU}(N_f)_R$ symmetry, just as the quark sector of the Standard Model, is a testing ground for the interplay of  quantum gravity with chiral symmetry which prevents a Dirac mass term. Spontaneous breaking of chiral symmetry would result in the generation of masses and the formation of bound states, analogous to the consequences of chiral symmetry breaking in QCD. In QCD, the masses are tied to the QCD scale, $\Lambda_{\rm QCD}$. Analogously, a quantum gravitational form of spontaneous chiral symmetry breaking would generically be tied to the Planck scale, resulting in the absence of light fermions. One can test for quantum-gravity induced chiral symmetry breaking by exploring whether four-fermion interactions remain finite under the impact of quantum gravity. These can be mapped into mass terms for bound states by a Hubbard Stratonovitch transformation, schematically $\lambda_4(\bar{\psi}\psi)^2 \rightarrow \frac{1}{\lambda_4}\phi^2 + \phi \bar{\psi}\psi$. Spontaneous symmetry breaking is signalled by the mass term transitioning from positive to negative values, and therefore tied to a divergence in the four-fermion coupling. Indeed, results in truncations indicate the existence of a (shifted Gaussian) fixed point for four-fermion couplings under the impact of quantum gravity, since the beta functions for the two four-fermion couplings $\lambda_{\pm}$ that form a Fierz-complete basis for the system read, see \cite{Eichhorn:2011pc}
\be
\beta_{\lambda_+}\!= 2 \lambda_+ + f_{\lambda_{\pm}}\lambda_++ \frac{5}{8} \frac{G^2}{(1-2\lambda)^3}+ \frac{3\lambda_+^2+2(1+N_f)\lambda_-\lambda_+}{8\pi^2}, \quad 
\beta_{\lambda_-}\!=2 \lambda_- + f_{\lambda_{\pm}}\lambda_-- \frac{5}{8} \frac{G^2}{(1-2\lambda)^3}+ \frac{(N_f-1)\lambda_-^2+N_f^2\lambda_+^2}{8\pi^2},\label{eq:betalambdapm}
\ee
where $f_{\lambda_{\pm}}$ is a gravity-induced anomalous dimension. The key term that shifts the free fixed point to an interacting one is the third term in each of the beta functions and has been written out explicitly in the Einstein-Hilbert truncation, for Landau-DeWitt gauge and a Litim-type cutoff \cite{Litim:2001up}. An inspection of Eq.~\eqref{eq:betalambdapm} reveals the existence of the shifted Gaussian fixed point for all positive values of $G$ and any\footnote{The restriction on the $\lambda$-interval arises since the flow equation \eqref{eq:floweq} depends on $\Gamma_k^{(2)}$, which features a cosmological-constant term. A calculation of gravitational fixed-point values reveals a fixed point within the interval $\lambda \in (-\infty, 1/2)$.} $\lambda \in (-\infty, 1/2)$, enabling the existence of light fermions and preserving the power-counting irrelevant nature of four-fermion interactions.\\
Beyond this fluctuation-induced mechanism for chiral symmetry breaking, there is classical symmetry-breaking through the background geometry, called gravitational catalysis, which occurs on spaces with negative curvature. The interplay of these two mechanisms has been explored for the first time in \cite{Gies:2018jnv}.

\subsubsection{Global symmetries}
The fermion system is also a useful testing ground for the fate of global symmetries in asymptotically safe gravity. In nongravitational systems, it holds that any global symmetry that is respected by the regulator will define a closed hypersurface under the RG flow. In other words, it is self-consistent to set all symmetry-breaking couplings to zero, as the RG flow does not generate them from couplings which respect the symmetry. In gravity, there are arguments suggesting that global symmetries might be broken explicitly by quantum effects, see, e.g., \cite{Kamionkowski:1992mf,Kallosh:1995hi}. Yet, within none of the truncations that have been explored, are there any indications for a global-symmetry breaking through quantum-gravity effects in various matter systems with various global symmetries, see \cite{Eichhorn:2017eht} for a discussion. 

Specifically, in \cite{Eichhorn:2011pc}, there is no explicit breaking of a global ${\rm SU}(N_f)_L \times {\rm SU}(N_f)_R$ symmetry, i.e., the only induced four-fermion interactions are those which respect this symmetry. In \cite{Eichhorn:2016vvy}, a Dirac mass term and a nonminimal curvature term are explicitly included in the truncation, i.e., the choice of truncation breaks this chiral symmetry. The results clearly show that the hypersurface with ${\rm SU}(N_f)_L \times {\rm SU}(N_f)_R$ symmetry is a closed hypersurface under the flow within the truncation; there are no contributions that generate the symmetry-breaking mass term and non-minimal coupling, if those are set to zero at some scale, see also \cite{deBrito:2019epw} for Majorana mass terms. \cite{Eichhorn:2016esv,Eichhorn:2017eht} explore fermion-scalar interactions in a system with $N_f=1$, and shows that derivative interactions which respect the chiral symmetry of the kinetic term are induced, i.e., cannot be consistently set to zero under the impact of quantum gravity. Yet, a symmetry-breaking Yukawa coupling can consistently be set to zero. Similar statements regarding the fate of global symmetries hold for other types of matter fields.

Moreover, there is a structural argument for the observed preservation of global symmetries in the flow-equation studies to date: The global symmetries of a given matter model are all exhibited by the kinetic terms of the various matter models -- otherwise they could not be global symmetries of the matter model without gravity. Now one asks whether it is consistent to set interactions to zero which would break these symmetries. This would \emph{not} be the case, if one could construct a diagram\footnote{The flow equation \eqref{eq:floweq} can be written as a sum of one-loop diagrams with an increasing number of vertices, by decomposing $\Gamma_k^{(2)}+R_k = \mathcal{P}_k + \mathcal{F}_k$, where $\mathcal{P}_k$ is the part that is independent of the dynamical fields and $\mathcal{F}_k$ carries the dependence on the dynamical field. This yields $\partial_t \Gamma_k = \frac{1}{2}{\rm Tr}\tilde{\partial}_t\ln \mathcal{P}_k+ \frac{1}{2}\sum_{n=1}^{\infty}\frac{(-1)^{n-1}}{n}{\rm Tr}\, \tilde{\partial}_t \left(\mathcal{P}_k^{-1}\mathcal{F}_k \right)^n$, where $\tilde{\partial}_t= \int \partial_t R_k \frac{\delta}{\delta R_k}$ acts only on the $k$-dependence in the regulator. Inspecting this form of the flow equation, one sees that each order $n$ in the $\mathcal{P}\mathcal{F}$ expansion corresponds to the sum of one-loop diagrams with $n$ vertices.} 
which would generate a symmetry-breaking interaction. Yet, the vertices in such a diagram would all have to arise from the kinetic term or other symmetric interactions. Therefore, the vertices transform in a well-defined representation of the global symmetry group (and not just a smaller symmetry group). The propagators in the diagram are either matter propagators (which again transform in a well-defined representation of the global symmetry group unless the regulator breaks it explicitly) or gravity propagators, which are \emph{blind} to internal symmetries, i.e., cannot carry a nontrivial representation of a smaller symmetry group. The trace in the flow equation then results in an invariant under the global symmetry group.
Thus, it is not possible to construct a diagram that would break the global symmetry and induce a flow of symmetry-breaking couplings, if these are set to zero at some scale.

The explicit studies mentioned above are all done in a Euclidean setting, and the arguments for quantum-gravity induced global symmetry breaking typically rely on evaporating black holes. Thus the final state of black-hole evaporation, which could depend on the quantum-gravity model under consideration \cite{Bonanno:2006eu}, matters for this argument. 
One might nevertheless wonder whether the preservation of global symmetries in the above studies could be an artefact of the Euclidean setup. Yet, the above argument about the construction of diagrams does not refer to a specific signature. Therefore, one would not expect  the result to  change in Lorentzian signature. Nevertheless, it would of course be interesting to check this explicitly, and evaluate RG flows where black-hole configurations are explicitly included in the metric configurations that are being integrated out. In particular, the effect of instanton-configuration as those explored in \cite{Lee:1988ge,Abbott:1989jw} has not been explicitly explored in the functional RG context.

\subsubsection{Upper bounds and predictions of couplings in Standard-Model like systems}\label{sec:upperbounds}
There are two effects of asymptotically safe gravity on Standard-Model like matter systems that have been found in truncated FRG flows. Both play a key role when attempting to understand whether the asymptotic-safety paradigm might provide a predictive ultraviolet completion of gravity-matter systems with a Standard-Model like matter content. In turn, both effects can provide nontrivial observational tests of asymptotically safe gravity.
\begin{itemize}
\item Under the impact of gravity, the free fixed point in a set (determined by the symmetry of the kinetic terms) of {\bf higher-order matter couplings} is shifted to a finite value, the {\bf shifted Gaussian fixed point} \cite{Eichhorn:2011pc,Eichhorn:2012va,Eichhorn:2017eht}. At this fixed point, these higher-order couplings are irrelevant, as follows from their canonical dimension. Both aspects can be elucidated on the example of the beta function for the coupling $w_2$ of the operator $(F_{\mu\nu}F^{\mu\nu})^2$ in an Abelian gauge theory, which reads \cite{Christiansen:2017gtg}
\be
\beta_{w_2} = 4 w_2 + 8 G^2- \frac{7}{2\pi}G\, w_2+ \frac{1}{8\pi^2}w_2^2,\label{eq:betaw2}
\ee
in a simple truncation that includes just the Newton coupling. While this is of course a very simple approximation, it suffices to show the existence of the shifted Gaussian fixed point, i.e., there is a fixed point at $w_{2\, \ast} \sim G$ (shifted away from $w_2=$ by gravitational effects), at which $w_2$ is irrelevant. Below the Planck scale, where $G \sim k^2$ quickly becomes tiny, the gravitational contribution vanishes very quickly, and the canonical scaling of the higher-order matter couplings implies that they go to zero in a power-law like fashion, and do therefore not leave a sizeable impact on $\Gamma_{k \rightarrow 0}$. The beta functions for the four-fermion couplings \eqref{eq:betalambdapm} are another paradigmatic example.\\ 
Note that for a subset of these couplings, such as for instance $w_2$, but not for example $\lambda_{\pm}$, the shifted Gaussian fixed point is shifted into the complex coupling plane if the gravity-contribution to the beta function is too large, see Eq.~\eqref{eq:betaw2}. This results in a bound on the gravitational fixed-point values, the {\bf weak-gravity bound} \cite{Eichhorn:2017eht}. The latter constraint on the gravitational fixed-point values arises from the simple observation that certain matter fields exist. If they cannot be included with a real-valued fixed point, the assumption of asymptotic safety is incompatible with the presence of these matter fields.
\item Under the impact of gravity, an anomalous scaling term is generated in the beta functions of {\bf canonically marginal couplings}. This term can balance against the non-gravitational contribution to generate an interacting fixed point, if the gravitational contribution features the appropriate sign. Crucially, if the coupling is marginally irrelevant without quantum gravity, it will be irrelevant at the interacting fixed point (and relevant at the free fixed point). This results in a unique Planck-scale value of the coupling, constituting an example of a prediction arising from the irrelevant direction of an interacting fixed point. Below the Planck scale, the logarithmic scale dependence of the coupling results in a ``memory" of Planck-scale physics present at IR scales, and translates into a prediction of the value of the coupling in the IR. \\
The interacting fixed point is necessarily accompanied by a free fixed point at which the coupling is relevant, as one can see by inspecting the beta function
\be
\beta_{g_i} = - f_{g_i}\, g_i + \beta_{g_i}^{(1)}g_i^3+...,
\ee
where $\beta_{g_i}^{(1)}$ is the (universal) one-loop term and $f_g$ parameterizes the (nonuniversal) gravity contribution in the FRG framework. Therefore, there are two distinct UV completions, both of which are interacting in the gravitational couplings, but one of which is also interacting in the matter sector, with the other one being free in the matter sector. The free fixed point is IR repulsive in $g_i$. Therefore, cross-over trajectories emanate from it and run towards the interacting fixed point. There are indications for such a fixed-point structure in the Abelian gauge coupling \cite{Harst:2011zx,Eichhorn:2017lry}, as well as Yukawa couplings \cite{Eichhorn:2017ylw,Eichhorn:2018whv}. For the Higgs quartic coupling, the situation is slightly different, with an infrared attractive fixed point at zero \cite{Shaposhnikov:2009pv} that is shifted to finite values if Yukawa couplings and the Abelian gauge coupling take finite fixed-point values \cite{Eichhorn:2017ylw}.
\end{itemize}

A combination of both effects has been used to propose in \cite{Eichhorn:2019yzm} that the predictive power of asymptotic safety could extend to parameters of the geometry, such as the dimensionality of spacetime. In a nutshell, the argument uses that the gravitational effects have to become large in $d>4$ in order for the fixed point in the Abelian gauge coupling to exist, as $f_g$ has to compete with an explicit dimensional scaling term in $d>4$. The second part of the argument uses the weak-gravity bound to suggest that a gravitational solution to the Landau-pole problem in the Abelian gauge coupling can only be achieved in $d>4$ at the expense of violating the weak-gravity bound and generating non-real-valued fixed points in higher-order couplings. Phrased differently, $d=4$ could be the only dimensionality, in which a near-perturbative UV completion of the Standard Model plus gravity at an asymptotically safe fixed point could be achievable. It goes without saying that the applicability of this argument to the Lorentzian setting, and its explicit tests in higher-order truncations are key outstanding tasks.

In all of the above, it is important to distinguish between universal and non-universal statements: The gravitational contribution to the flow of marginal matter couplings has non-universal parts as well as universal parts (essentially those proportional to dimensionless gravitational couplings (i.e., curvature squared couplings) or dimensionless combinations of gravitational couplings, e.g., the product $G \lambda$). The beta functions for higher-order matter couplings are non-universal, both in their gravitational as well as non-gravitational parts. In general, non-universality affects all beta functions (and therefore running couplings), however, for marginal couplings it only sets in at three loops in massless schemes, which is why the leading terms in the Standard Model beta functions without gravity are universal. Therefore, the actual flows can look quite differently in different schemes, and it is in fact possible to set certain contributions to zero, e.g., by a choice of regulator, see, e.g., \cite{Folkerts:2011jz}. The key point to note is that the existence of a fixed point is a universal statement. Secondly, critical exponents are universal. Therefore the number of free parameters of the system does not depend on the choice of regulator, and the number of relations imposed between various couplings do not change. Accordingly, while flows might differ, the statement that the IR values of various couplings are subject to relations due to the existence of irrelevant directions at the fixed point, is general. If the number of relevant directions is smaller than the number of renormalizable matter couplings plus measured gravitational couplings, then the asymptotically safe scenario is testable.\\
 The key challenge then becomes to calculate the IR values with sufficient precision, requiring a good control of truncations. From a purely technical point, one should note that scales $k\geq 10^{19}\, \rm GeV$ are dominated by fixed-point behavior. At $k\lesssim 10^{19}\, \rm GeV$, gravity fluctuations are quickly driven to zero, leaving the RG flow of a pure matter model. Within the functional RG, the nontrivial denominator of the flow equation encodes threshold effects which lead to automatic decoupling of massive modes once $k$ falls below the corresponding mass scale.

\subsubsection{Outlook: Learning about dark sectors}
The very first steps to explore the interplay of asymptotically safe gravity with dark matter models have been taken in \cite{Eichhorn:2017als,Reichert:2019car}, see also \cite{Kwapisz:2019wrl} for another BSM setting. A general remark is that one might expect asymptotic safety to reduce the parameter space for dark-matter searches, since a dark-matter coupling that is a free parameter in an EFT setting could become an irrelevant coupling at an asymptotically safe fixed point, as tentatively also explored in a setting without gravity \cite{Eichhorn:2018vah}. Such constraints, together with relic-density constraints and constraints from direct searches might even allow to rule out dark-matter candidates in an asymptotically safe context.

Furthermore, indirect constraints could arise on the number of fields in a dark sector: Even if a dark sector is purely gravitationally coupled, it might be constrained observationally within the asymptotic-safety framework, as discussed in \cite{Eichhorn:2017ylw}: The key idea here is to use the upper bounds on SM couplings as discussed in Sec.~\ref{sec:upperbounds}. For this argument, we will assume that the fixed-point structure discussed above, for which tentative hints exist in simple truncations, indeed persists.
Then, the values of the upper bounds depend on the gravitational fixed-point values. In turn, these depend on the number and type of matter fields. Accordingly, changing the number of purely gravitationally coupled, i.e., ``dark", matter degrees of freedom changes the values of the upper bounds. If these fall below the experimentally determined values of the couplings, the model is ruled out observationally.

\section{Relation to other approaches to quantum gravity and conclusion}\label{sec:relntoapproaches}

In general one would not expect that without much experimental guidance, i.e.,  on purely theoretical grounds, we can figure out a viable description of quantum gravity. Therefore, asymptotic safety  (or any of the other quantum-gravity approaches being explored) is most likely not the full story of quantum gravity. Rather, the various approaches explore various aspects that might be important for a quantum theory of gravity. Hence each approach could have something important to teach us about a full description of quantum spacetime. In this view of the current state of research on quantum gravity, it is important to search for potential connections between approaches. \\
To highlight that asymptotically safe gravity is not a priori mutually exclusive with other approaches to quantum gravity, let us briefly review the idea of effective asymptotic safety. Here, the emphasis is not on the possibility to construct a theory viable up to arbitrarily high scales. Instead, the emphasis is on the predictivity that the existence of a fixed point entails: As highlighted in Fig.~\ref{fig:illFP}, trajectories that start at some point in the space of couplings at some cutoff scale $k'$, can be pulled towards the RG fixed point and spend a large amount of RG ``time" (i.e., a large range of scales) close to the fixed point, in a very nearly scale-invariant regime. Towards the IR, those trajectories then stay close to the critical surface, and result in IR predictions very similar to those on an actually asymptotically safe trajectory. Therefore, asymptotic safety might turn out to be an intermediate description of quantum gravity, superseded at some very high scale by a more microscopic description, while making the microscopic description more predictive, see \cite{Percacci:2010af} for the general idea and \cite{deAlwis:2019aud} for a discussion in the context of string theory.

Besides the conceptual basis, the methods used to explore the various quantum-gravity approaches also tend to differ, and there is a case to be made that the application of various methods across different approaches could be important to achieve progress. In this context, let us highlight that Renormalization Group techniques are now being used in a number of quantum gravity approaches, see, e.g., \cite{Rivasseau:2011hm,Delcamp:2016dqo,Carrozza:2016vsq,Bahr:2018gwf,Eichhorn:2018phj,Ambjorn:2020rcn} and references therein. The key point in these cases is the search for universality in the continuum limit, which ensures that the physics that is being described in these approaches is independent of unphysical microscopic details (such as, e.g., a particular choice of discretization), and that continuum symmetries such as diffeomorphism symmetry, are respected.  \newline

In summary, asymptotically safe gravity is a conservative approach to quantum gravity which has seen considerable progress over the last years, with indications for the Reuter fixed point in Euclidean quantum gravity becoming more and more compelling. Important questions on the technical and conceptual side are currently open, while first steps towards answering many of them are being made. In particular, it is an encouraging perspective that ruling out this approach to quantum gravity observationally appears to be an actual possibility without the need to explore Planck-scale physics directly. The wealth of available data on particle physics can in principle be used to constrain any quantum-gravity approach, as it is a nontrivial requirement that the observed IR physics can be obtained as the effective large-scale description of a given microscopic model including quantum gravity. As examples, we have discussed the number of various matter fields, the lightness of fermions, the various interaction strengths of Standard Model matter fields and more. In practise, making use of these observational tests is an outstanding challenge in most quantum gravity approaches, since it is in many settings immensely difficult to calculate the consequences of a given model for the matter sector. In asymptotically safe gravity, proofs-of-principle have been provided that this is possible, making it a viable possibility to rule out this idea with the help of already available experimental data.\newline\\

\emph{Acknowledgements:} I acknowledge support by a research grant (29405) from VILLUM FONDEN, the DFG (Deutsche Forschungsgemeinschaft) under grant no Ei/1037-1, and the Perimeter Institute for Theoretical Physics under the Visiting Fellowship program.

\end{document}